\documentclass[%
reprint,
amsmath,
amssymb,
aps,
]{revtex4-2}

\usepackage{graphicx}
\usepackage{placeins}
\graphicspath{{figs/}}

\begin{document}

\title{Edge Currents Shape Condensates in Chiral Active Matter}

\author{Boyi Wang$^{1,2}$}
\email{boyiw@pks.mpg.de}
\author{Patrick Pietzonka$^{3}$}
\email{p.pietzonka@ed.ac.uk}
\author{Frank J\"ulicher$^{1,2,4}$}
\email{julicher@pks.mpg.de}

\affiliation{%
$^1$Max Planck Institute for the Physics of Complex Systems, 01187 Dresden, Germany}
\affiliation{%
$^2$Center for Systems Biology Dresden, Pfotenhauerstrasse 108, 01307 Dresden, Germany}
\affiliation{%
$^3$SUPA, School of Physics and Astronomy, University of Edinburgh, Peter Guthrie Tait Road, Edinburgh EH9 3FD, United Kingdom}
\affiliation{%
$^4$Cluster of Excellence Physics of Life, TU Dresden, 01062 Dresden, Germany}

\begin{abstract}

Chiral active matter, which breaks both parity symmetry and time-reversal symmetry, is ubiquitous in living systems. Here, we introduce a minimal two-dimensional chiral active lattice gas by incorporating stochastic, biased local rotations. 
At low temperatures, the system coarsens into condensates with chiral orientations and faceted, crystal-like shapes. 
The interfaces align at characteristic angles with respect to the lattice axes and exhibit edge currents that are persistent, unidirectional, and angle-dependent. 
To generalize these findings, we propose a continuum theory by adding an active chiral edge current term to Model B, which reveals the essential role of active chiral transport in the interfacial dynamics of phase separation. 
Edge currents with $n$-fold symmetry produce condensates whose shapes resemble regular $n$-sided polygons. 
In the thin-interface limit, we construct an effective interface potential governing edge currents, from which the steady-state condensate geometry can be obtained, both in the lattice model and the continuum description.

\end{abstract}
\maketitle

\section{Introduction}

Chirality, the absence of mirror symmetry in an object or process, is a ubiquitous feature of nature~\cite{prelog1976chirality,harris1999molecular,barron2008chirality}. 
Chiral systems can be either passive or active. Active chiral systems lack mirror symmetry and simultaneously break time-reversal symmetry through sustained energy input~\cite{sase1997axial,tsai2005a,liebchen2022chiral}.
In biological systems, chiral active matter plays important roles in many active cellular processes.
Cells rotating near interfaces generate chiral flows and nonequilibrium patterns~\cite{tan2022odd,espadaburriel2024active} and multicellular assemblies such as spherical organoids can spontaneously rotate in a three-dimensional matrix exhibiting spontaneous chirality~\cite{tan2024emergent}.
Across scales, active torques generated by molecular motors and cytoskeletal filaments~\cite{sase1997axial,shankar2024active} can be amplified into cellular chirality~\cite{tee2015cellular} and eventually to tissue-scale chiral dynamics~\cite{chen2025chirality} and embryonic left-right symmetry breaking~\cite{hamada2002establishment,raya2006left,hirokawa2006nodal,naganathan2014active}.
Behaviors of chiral active matter have been studied in artificial systems, both experimentally and using simulations. Examples include active spinners, chiral active Brownian particles, and active chiral granular systems. These systems exhibit a rich phenomenology, including boundary-induced oscillatory collective dynamics~\cite{tsai2005a,vanzuiden2016spatiotemporal,liu2020oscillating,kole2024chirality}, odd rheological responses~\cite{banerjee2017odd,scheibner2020odd,kole2021layered}, phase separation with edge currents~\cite{soni2019the,zhao2021emergent,langford2025phase,caprini2025bubble} and persistent boundary transport associated with lift forces~\cite{caprini2019active,yang2021topologically,lou2022odd,hosaka2021hydrodynamic,lier2023lift,kant2025edge}. Additional nonequilibrium features such as glassy dynamics~\cite{debets2023glassy}, hyperuniform states~\cite{zhang2022hyperuniform,wang2025hyperuniform}, synchronization~\cite{uchida2010synchronization,samatas2023hydrodynamic,xia2024biomimetic} and spontaneously rotating droplets~\cite{carenza2019rotation,wang2021spontaneous,nejad2023spontaneous} have also been reported. 

Motivated by these phenomena, particle-based, rheological and hydrodynamic theoretical frameworks have been developed~\cite{furthauer2012active,furthauer2013active,banerjee2021active,han2021fluctuating,markovich2021odd,sese2022microscopic,kreienkamp2022clustering,ishimoto2023odd,fruchart2023odd,hosaka2023hydrodynamics, markovich2024nonreciprocity,ding2024odd,kalz2024field,markovich2025chiral}.
While these theories describe chiral active matter through gradient expansions around a homogeneous nonequilibrium state, a complementary approach in statistical physics is to use lattice models to study spatial patterns and their nonequilibrium dynamics~\cite{huggins1941solutions,flory1942thermodynamics,grinstein1985statistical,kawamura1992chiral,rothman1994lattice,odor2004universality,marro2005nonequilibrium, solon2013revisiting,ccauglar2016chiral,houssene2018exact,weber2019physics,metson2025continuous}. Among these, the lattice gas (or equivalently the Ising model) constitutes a minimal microscopic description of binary systems, allowing symmetry-breaking effects to be introduced, such as activity~\cite{solon2015flocking,agranov2022entropy,scandolo2023active}, chirality~\cite{wang2024condensate,de2025simple} and non-reciprocity~\cite{avni2025nonreciprocal,blom2025local,blom2026dynamic}. 
Moreover, for number conserving kinetics, the lattice gas admits a systematic coarse-graining to Model-B--type field theories~\cite{kawasaki1966diffusion,hohenberg1977theory}, whose active extensions have been used to describe motility-induced phase separation in active systems~\cite{wittkowski2014scalar,cates2015motility,tjhung2018cluster}.

In this work, we propose a minimal chiral active lattice gas in two dimensions, in which chirality and activity are introduced through stochastic biased local rotations. 
We study phase separation and collective behaviors that emerge in conserved binary systems that are both chiral and active.
Numerical simulations reveal a rich phenomenology, indicating that microscopic chiral dynamics can reorganize phase-separated states at the macroscopic level. 
To provide a theoretical description, we develop a coarse-grained continuum theory by extending Model-B dynamics to include a chiral active transport term. 
Within this framework, we define an effective interface potential and construct a dynamic principle for determining the fluctuating steady-state geometry of condensates, phase-separated domains enriched in one component.

\section{\label{sec:lattice} minimal chiral active lattice gas}
\begin{figure}[b]
\includegraphics[width=0.47\textwidth]{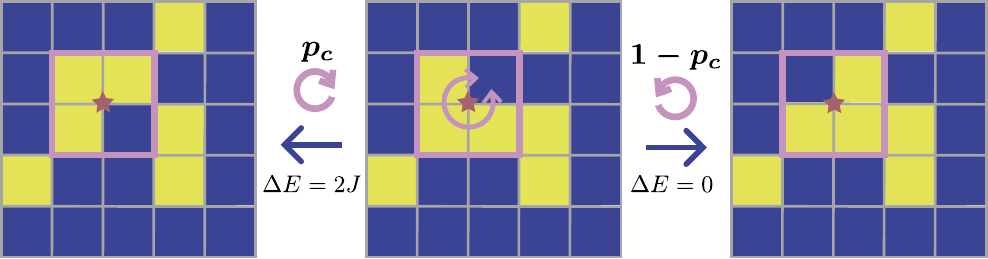}
\caption{\label{fig:rotation} Elementary move of the chiral active lattice gas. Starting from the present configuration (middle), a square of $2\times 2$ lattice sites is randomly selected for rotation in counterclockwise direction at probability $1-p_c$ (left) or in clockwise direction at probability $p_c$ (right). Acceptance probabilities are determined by the change of Ising energy.}
\end{figure}

We consider a two-dimensional square lattice with $N$ sites and periodic boundary conditions, where each site $\mathbf{m} = (i,j)$ hosts an Ising spin $\sigma_{\mathbf{m}} = \pm 1$. 
As a lattice gas, these two states represent two different types of particles (A and B), where the total number of A is $N_A=\sum_\mathbf{m} \delta_{\sigma_{\mathbf{m}},1}$ and the total number of B is $N_B=\sum_\mathbf{m} \delta_{\sigma_{\mathbf{m}},-1}$, with $N=N_A+N_B$. 
With number conservation, $N_A$ and $N_B$ are fixed. 
For every microstate $\{\sigma_{\mathbf{m}}\}$ of the lattice field, the Ising energy reads
\begin{equation}
E(\{\sigma_{\mathbf{m}}\}) = -J \sum_{\langle \mathbf{m},\mathbf{n}\rangle} \sigma_{\mathbf{m}} \sigma_{\mathbf{n}},
\label{eq:energy}
\end{equation}
where $J>0$ is the coupling constant for nearest-neighbor interaction and the sum runs over nearest-neighbor pairs $\langle \mathbf{m},\mathbf{n}\rangle$. 

We introduce a minimal dynamics to break both time-reversal symmetry and parity symmetry through stochastic biased local rotations of this lattice gas.
At each update attempt, a $2\times2$ box of lattice sites is randomly selected, as indicated by the pink box in Fig.~\ref{fig:rotation}.
With the box center fixed, this box can be rotated by $\pi/2$ in either a clockwise or counterclockwise direction. The probability of attempting a clockwise rotation is $p_c$, while that of a counterclockwise rotation is $1-p_c$. This attempted rotation would change the  microstate $\{\sigma_{\boldsymbol{m}} \}$ to $\{\sigma'_{\boldsymbol{m}} \}$ and be accepted according to the Metropolis criterion~\cite{hastings1970monte} with the acceptance probability: 
\begin{equation}
    P_\text{acc}(\Delta E) = \min(1, e^{-\frac{\Delta E}{k_B T}}),
\end{equation}
where $\Delta E = E(\{ \sigma'_{\boldsymbol{m}} \}) - E(\{\sigma_{\boldsymbol{m}}\})$ is the energy difference between the configurations before and after the local rotation, $k_B$ is the Boltzmann constant, and $T$ is temperature. 
The combined selection and acceptance probability of this rotation is therefore $P_\circlearrowright(\{ \sigma_{\boldsymbol{m}} \})=\frac{p_c}{N}P_\text{acc}(\Delta E)$ when it is in clockwise direction and $P_\circlearrowleft(\{ \sigma_{\boldsymbol{m}} \})=\frac{1-p_c}{N}P_\text{acc}(\Delta E)$ when it is in counterclockwise direction.
We can consider the bias of the rotation as a form of activity, which can be quantified by the free energy difference  $\Delta \mu=k_BT\ln[p_c/(1-p_c)]$ supplied by a fuel reservoir.
This leads to the local detailed balance relation
\begin{equation} 
\ln\frac{P_\circlearrowright(\{ \sigma_{\boldsymbol{m}} \})}{P_\circlearrowleft(\{ \sigma'_{\boldsymbol{m}} \})} = \frac{\Delta\mu-\Delta E}{k_BT}
\label{eq:e_drive} 
\end{equation}
between the probability of a clockwise rotation and the reverse counterclockwise rotation around the same rotation center. 
When $p_c = 1/2$, the driving is $\Delta \mu = 0$ such that Eq.~\eqref{eq:e_drive} becomes a global detailed balance relation. Then, the system is at equilibrium and not chiral, corresponding to an Ising model with number conservation. Indeed, one can show that the dynamics using only local rotations is ergodic, i.e., the system can reach any microstate, as proven in Appendix~\ref{app:ergodicity}. When $p_c \neq 1/2$, the system is chiral and maintained out of equilibrium, which implies that both parity and time reversal symmetry are broken. It is still ergodic, but the probability of a microstate is no longer determined by only its energy.

\section{\label{sec:current} emergent chiral currents along Faceted Interfaces}

\begin{figure}[b]
\includegraphics[width=0.47\textwidth]{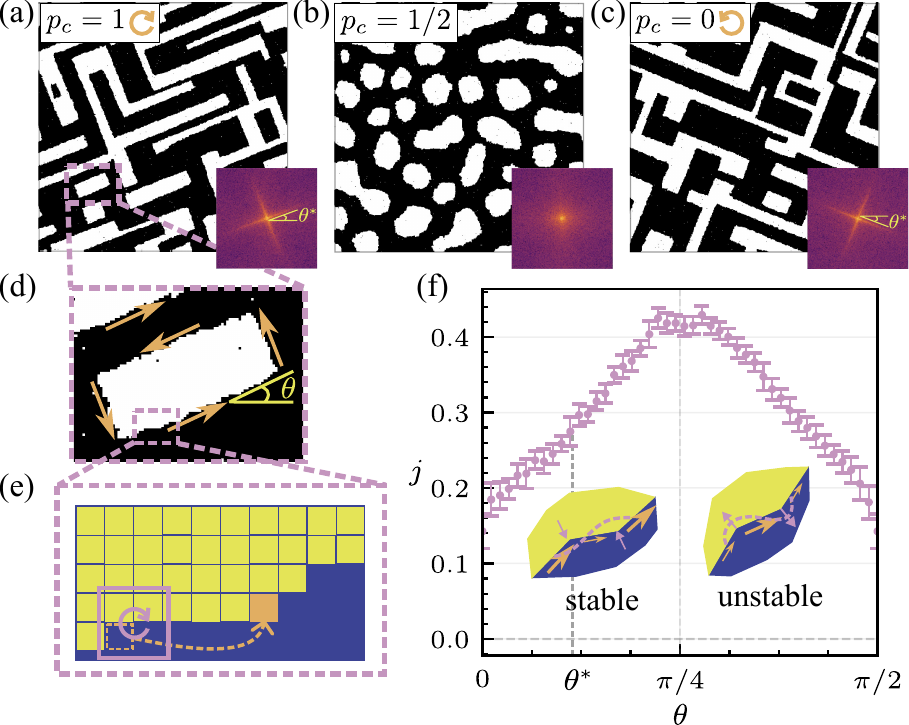}
\caption{\label{fig:simulation} 
Simulations of the chiral active lattice gas. Panels (a)-(c) show snapshots of lattice configurations for $k_BT/J=1.2$ on a $512\times512$ lattice after $10^6$ Monte Carlo sweeps, starting from random initial conditions with $40\%$ type-A (white) and $60\%$ type-B (black) particles.
(a) $p_c=1$, clockwise-only rotations;
(b) $p_c=1/2$, unbiased rotations;
(c) $p_c=0$, counterclockwise-only rotations.
Insets show the corresponding two-dimensional pattern Fourier spectra.
(d) Zoom-in of a rectangular condensate illustrating the direction of the  edge current at a straight interface with the tilt angle $\theta$. Panel (e) shows a microscopic interface between type~A (yellow) and type~B (blue) particles. The transport of a particle of type~A along the interface, via repeated rotations, contributes to the  edge current. Panel (f) shows the dependence of the edge-current magnitude $j$ on the interface angle $\theta$, measured from simulations of linear interfaces. Error bars indicate the standard deviation over 12 independent runs. Insets illustrate the stability (left) and instability (right) of interfaces depending on the sign of
$\mathrm{d}j/\mathrm{d}\theta$. Orange arrows indicate the current direction and magnitude of the yellow phase, which as a result evolves in the direction indicated by the pink arrows and dashed lines.
}  
\end{figure}

We study the phase separation behaviors using kinetic Monte Carlo simulations~\cite{frenkel2023understanding} with the dynamics defined above. 
One Monte Carlo sweep is defined as $L^2$ local rotation attempts on a square lattice of size $L\times L$, such that each $2\times2$ box is selected once per sweep on average. 
We set the simulation temperature with $k_BT/J=1.2$, well below the equilibrium square-lattice Ising critical temperature, for which $k_BT_c/J\simeq2.27$~\cite{onsager1944crystal} ($T\simeq0.53\,T_c$), ensuring strong phase separation while retaining fluctuations.
Figure~\ref{fig:simulation}a--c shows snapshots of lattice configurations obtained during coarsening for different rotation biases $p_c$, along with the corresponding pattern Fourier spectra~\cite{SM}.

Starting from random initial conditions, biased clockwise ($p_c=1$, Fig.~\ref{fig:simulation}a) and counterclockwise ($p_c=0$, Fig.~\ref{fig:simulation}c) rotations drive the system into nonequilibrium
maze-like condensates with approximately straight interfaces tilted at a characteristic angle $\theta^*$ with respect to the lattice axes. These two cases are mirror images of each other, reflecting the opposite handedness of the imposed chirality.
Since $\theta^*$ is not a multiple of $\pi/4$, the two mirror orientations cannot be superposed by a lattice symmetry operation and are therefore chiral. 
In contrast, for unbiased rotations ($p_c=1/2$, Fig.~\ref{fig:simulation}b), the system forms irregular condensates without any preferred interface orientation, similar to the condensate shape and orientation under equilibrium Kawasaki dynamics~\cite{kawasaki1972kinetics}. 
These emerging patterns (Fig.~\ref{fig:simulation}a-c) are further characterized by their two-dimensional Fourier spectra (insets in the bottom-right of each panel).
In the chiral cases, the spectra exhibit a tilted cross-shaped structure, indicating a pronounced fourfold orientational anisotropy. The dominant wavevectors are organized along four symmetry-related directions, with the smallest angle relative to the horizontal lattice axis given by
$\theta^* = 0.3854 \pm 0.0007$ (averaged over 32 parallel simulations), together with symmetric counterparts at
$\theta^* + n\pi/2$ with $n=0,1,2,3$. In Appendix~\ref{app:detail}, we provide a quantitative analysis of these simulations.

We next examine the particle transport induced by chirally biased local rotations. At low temperatures, the system exhibits strong phase segregation, with each domain containing almost only one type of particles. In the bulk of these domains, local rotations move a single particle in any direction at equal probability, such that its effective dynamics becomes diffusive and unbiased. 
In contrast, at the boundary of a condensate, local rotations give rise to persistent, unidirectional particle transport along the interface, i.e., an edge current (Fig.~\ref{fig:simulation}d). The microscopic origin of the  edge current is illustrated in Fig.~\ref{fig:simulation}e, with further details shown in Fig.~\ref{fig:element} in the appendix.
A tilted interface between two domains of the lattice has a staircase-like pattern. Only particles at the two ends of each step can easily be moved by a rotation, because such rotations incur a smaller change in Ising energy than a rotation in the middle of a step. 
After an initial rotation at one end of a step, subsequent local rotations occur very fast, since they incur no increase of Ising energy. These rotations transport particles from one end of a step to the next, through particles of type~A moving in one direction or particles of type~B (or ``holes'' from the perspective of the component~A) moving in the opposite direction. This dynamics results in a net tangential flux along the interface. 
The direction of this  edge current reverses under opposite rotation bias, reflecting that global chiral active transport emerges from local chiral dynamics.

We quantify this effect by defining an effective current magnitude $j(\theta)$ of particles of type~A transported along an edge per Monte-Carlo sweep. It can be measured in simulations with skewed periodic boundary conditions enforcing a single interface at prescribed tilt angles from $0$ to $\pi/2$ (Appendix~\ref{app:edge_current}). Figure~\ref{fig:simulation}f shows that $j(\theta)$ exhibits a pronounced angular dependence and reaches a maximum at $\theta \approx \pi/4$. Interfaces whose orientation lies in the regime where the current increases with $\theta$ are stable,
whereas interfaces in the opposite regime are unstable and reshape under the action of the current, through the amplification of perturbations.

\section{\label{sec:camb} Continuum model:\\ Chiral active Model~B}
\begin{figure}[b]
\includegraphics[width=1\linewidth]{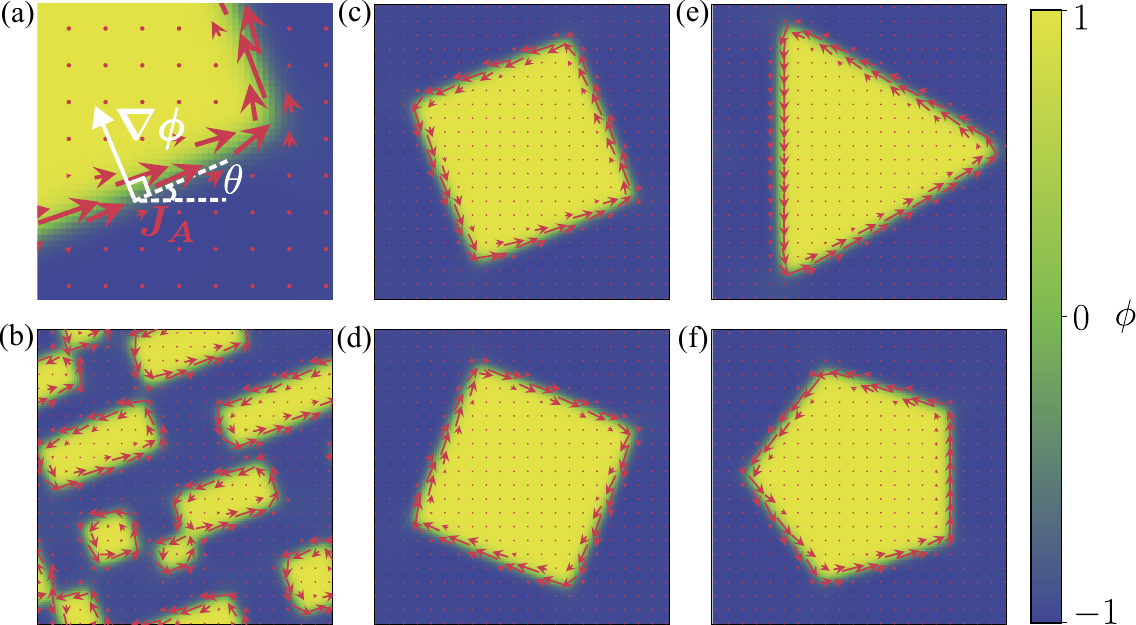}
\caption{Illustration and numerical solutions of the chiral active Model~B, with the two-dimensional scalar field $\phi$ indicated by the colourmap. (a) The active force is defined such that it drives an  edge current $\mathbf{J}_A$ (red) in a direction perpendicular to the gradient $\nabla\phi$, i.e., tangentially along the interface. The strength of the active force depends on the magnitude of $\nabla \phi$ and is modulated by the angle $\theta$ between the interface and the horizontal direction. (b) Snapshot of a field at time 8000 after a random start, with 4-fold symmetry and activity parameter $\zeta=0.1$. Panels (c) and (d) illustrate the chirality of the system, leading to mirrored orientations of stationary condensates for $\zeta=+0.1$ and $\zeta=-0.1$. Panels (e) and (f) show stationary condensates for $n=3$ and $n=5$ fold symmetry. Other parameters are $a=-1/4$, $b=1/4$, $M=1$, $K=1$, $j_0=3/2$ and $|\zeta|=0.1$. }
\label{fig:camb} 
\end{figure}

At equilibrium, phase separation in number-conserved Ising systems is well described by Model~B dynamics~\cite{kawasaki1966diffusion,hohenberg1977theory}, which predicts circular condensates in two dimensions, with geometry controlled solely by interfacial energy. 
Motivated by the observation of directional edge currents in our chiral active lattice gas, we introduce an active edge current as a minimal chiral nonequilibrium extension, which we refer to as chiral active Model~B. 

We consider a scalar order parameter $\phi(\mathbf r,t)$ representing a conserved local density field.
The passive properties of the system are described by the standard Ginzburg--Landau free-energy functional
\begin{equation}
F[\phi]
=\int \mathrm d\mathbf r\,
\left(
\frac a2\phi^2
+\frac b4\phi^4
+\frac K2(\nabla\phi)^2
\right)
\label{eq:freeenergy}
\end{equation}
with $a<0$ and $b>0$,
which gives rise to the chemical potential
\begin{equation}
\label{eq:mu}
\mu
=\frac{\delta F}{\delta\phi}
=a\phi+b\phi^3-K\nabla^2\phi .
\end{equation}
The dynamics of the conserved scalar field follows a Model~B (Cahn--Hilliard) form,
\begin{equation}
\label{eq:camb}
\left\{
\begin{aligned}
\partial_t\phi &= -\nabla\cdot\mathbf J,\\
\mathbf J &= -M\nabla\mu + \mathbf J_A,
\end{aligned}
\right.
\end{equation}
where $M$ is the effective mobility and we introduce the active edge current 
\begin{equation}
\label{eq:JA}
\mathbf J_A = j_A(\theta)\,\boldsymbol\varepsilon\,\cdot\nabla\phi.
\end{equation}
Here, $j_A(\theta)$ is a function of the oriented interface angle
$\theta\in[0,2\pi)$ and 
\begin{equation}
\boldsymbol\varepsilon =
\begin{pmatrix}
0 & 1\\
-1 & 0
\end{pmatrix}
\end{equation}
is a clockwise $\pi/2$ rotation matrix in two dimensions.
At points on an interface where
$|\nabla\phi|\neq0$, we define $\theta$ via
\begin{equation}
\label{eq:theta}
    (\cos\theta,\sin\theta)
    =
    \frac{\boldsymbol\varepsilon\cdot\nabla\phi}
         {|\nabla\phi|}
    .
\end{equation}
If the field has the form of domains of constant $\phi$, separated by smooth interfaces, the gradient is non-zero only at these interfaces.
Since $\boldsymbol\varepsilon\cdot\nabla\phi$ is orthogonal to $\nabla\phi$,
the active current $\mathbf J_A$ flows along interfaces at a strength that is determined by the local angle $\theta$ of the interface to the $x$-axis (Fig.~\ref{fig:camb}a).

Choosing an nonzero edge current $j_A$ that is independent of $\theta$ leads to a model whose density field $\phi$ evolves exactly as in the equilibrium Cahn--Hilliard model, while retaining chiral active transport along interfaces. This is because partial derivatives commute, so that $\nabla\cdot\boldsymbol{\varepsilon}\cdot\nabla=0$, which eliminates the active term in the expression for $\partial_t\phi$ in Eq.~\eqref{eq:camb}. In contrast, an effect of the nonequilibrium dynamics on $\phi$ can be expected whenever $j_A(\theta)$ is $\theta$-dependent, as it is also the case for the edge currents observed in our lattice model.
Motivated by the 4-fold-symmetric edge current in the chiral active lattice gas, we construct an $n$-fold-symmetric active edge current $\mathbf J_A$, with
\begin{equation}
     j_A(\theta) = \zeta (j_0 - \cos(n\theta))
\label{eq:cambja}
\end{equation}
as an example, conceptually the same edge current as the one measured in Fig.~\ref{fig:simulation}f when $n=4$. Here, $j_0$ is a constant, and $\zeta$ determines the direction and magnitude of the edge current induced by the interface slope. 

Numerical solutions of the chiral active Model~B with periodic boundary conditions (see Appendix~\ref{app:camb_numerics}) show the emergence of condensates with shapes that are characteristic for the symmetry of the  edge current.
Starting from a random initial configuration, a maze-like structure with a preferred orientation emerges during coarsening for a 4-fold-symmetric $\mathbf J_A$ (Fig.~\ref{fig:camb}b and Ref.~\cite{SM}), similar to what we observe for the chiral active lattice gas. After the coarsening (or after reshaping, if the system is initialized as a circular condensate), the system reaches a steady state with a single domain of $\phi\approx 1$, and a single domain of $\phi\approx -1$, as shown in Figs.~\ref{fig:camb}c-f and Ref.~\cite{SM}. 
The chirality of the system is exemplified in Figs.~\ref{fig:camb}c,d, showing the steady states for 4-fold symmetric $\mathbf J_A$ with opposite chirality, which leads to tilted square condensates that are mirror reflections of each other. Furthermore, we can also construct other forms of symmetry. 
The resulting edge-currents with 3-fold and 5-fold symmetry produce corresponding $n$-sided polygons, as shown in Fig.~\ref{fig:camb}e,f. 
All these faceted interfaces are a clear departure from the circular condensates that are the steady state of the equilibrium Cahn--Hilliard model, which are only recovered for the angle-independent constant edge current~($n=0$). 
The numerical solutions also confirm that introducing the $\mathbf J_A$ does not modify the bulk $\phi$ values nor the interfacial profile distinctively compared to the Cahn--Hilliard model (Appendix~\ref{app:profile}).
The observed polygonal shape of condensates and angular orientation of their edges therefore arise purely from the chiral dynamics at interfaces. A linear stability analysis of steady straight interfaces in the thin-interface limit is presented in Appendix~\ref{app:stability}, where we show that the angle dependent edge current enters the real part of the interface growth rate and controls long-wavelength stability.

\begin{figure}[b]
\includegraphics[width=1\linewidth]{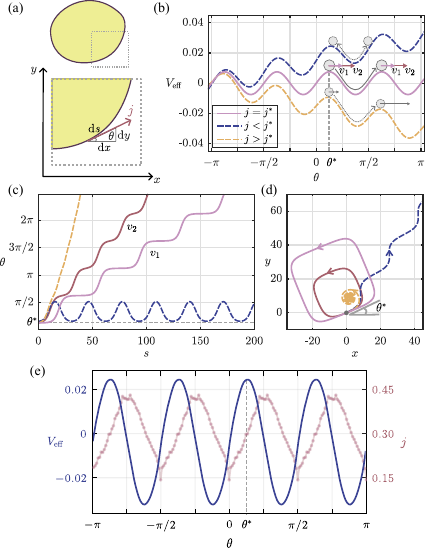}
\caption{
Construction of condensate shapes and orientations in the thin interface limit. (a) The interface between the inside (yellow) and outside (white) of the condensate is parameterized by the coordinates $x(s)$ and $y(s)$, with an edge current $j$ flowing along the tangential direction at angle $\theta$. 
(b) Effective potential $V_\text{eff}$ corresponding to Eq.~\eqref{eq:jA_4fold} (with $j_0=1$, $\zeta=0.03$, $\beta=1$) leading to the dynamics of a particle in the Newton mapping (gray ball). The overall tilt of the potential is determined by $j$ (yellow: 0.02, pink: 0.03, blue: 0.04) in relation to the steady current $j^*=0.03$. Arrows indicate the particle's velocity in our examples (considering two different velocities $v_1$ and $v_2$ for the scenario with $j=j^*$). The locus $\theta^*$ of a potential maximum, where the particle moves most slowly, determines a stable slope of the condensate contour.
(c) Trajectories $\theta(s)$ of a particle in the Newton mapping for the scenarios above.
(d) Condensate contours corresponding to the trajectories above. Only the potential with $j=j^*$ leads to closed contours, otherwise one obtains an inspiral (yellow), or an extended wavy interface (blue). (e) Effective potential (blue) constructed from the current-angle relation (red) obtained from simulations of the lattice model. The maximum of the potential at $\theta^* \approx 0.384$ matches with the stationary angle $\theta^* = 0.3854 \pm 0.0007$ of interfaces observed in simulations.}
\label{fig:newton} 
\end{figure}

\section{Effective interface potential and condensate geometry}
\label{sec:eff_potential_geometry}

In equilibrium, faceted condensates are a result of anisotropic surface tension. A classic example is provided by crystals, whose shape can be determined via the Wulff construction~\cite{wulff1901question}. Surface tension anisotropy also shapes condensates in multicomponent continuum models~\cite{jeong2014chiral,haas2017theory,haas2019shape,garcia2021faceting}. 
However, out of equilibrium, the steady interface shape is generally not determined by minimization of a free-energy functional.
Instead, we here define an effective interface potential that governs edge currents and develop a dynamic principle that determines the geometry and orientation of condensates with active edge currents. It is applicable to both continuum and lattice descriptions.

We consider condensates which are much larger than the typical width of their interfaces. In this thin interface limit, a condensate can be represented by the shape of its outline in the $x$--$y$ plane (Fig.~\ref{fig:newton}a). 
We write this outline as $\mathbf{r}(s)=(x(s),y(s))$, using the arclength $s$ from an arbitrary reference point $\mathbf{r}(0)$ on the interface as a parameter. 
We use the convention that $s$ increases in counterclockwise direction when $\phi$ is bigger inside the condensate.
The unit tangent vector ${\bf t}(s)=\mathrm{d}{\bf r}/\mathrm{d}s$ is parameterized by the local orientation angle $\theta(s)$ of the interface as
\begin{equation}
{\bf t}(s)=(\cos\theta(s),\sin\theta(s)).
\end{equation}
With these conventions, the local curvature of the interface is
\begin{equation}
\kappa(s) = \frac{ \mathrm{d}\theta}{\mathrm{d}s}.
\end{equation}

Given that the interfacial density profile remains close to the equilibrium Model~B and the interface-averaged chemical potential
of circular condensates varies approximately linearly with curvature (Appendix~\ref{app:profile}), we adopt the equilibrium thin-interface approximation.
Within this approximation, the Gibbs-Thomson relation~\cite{bray1994theory} determines the local chemical potential along the interface as 
\begin{equation}
  \mu(s) = \mu_0 + \beta\,\kappa(s),
\label{eq:GibbsThomson}
\end{equation}
where $\mu_0$ is the chemical potential of a straight interface and $\beta$ is the effective capillary tension.

In the steady state, bulk fluxes vanish and mass transport is confined to the interface, without sources or sinks.
We thus describe the tangential transport by an effective one-dimensional current along the interface, written in nondimensional units in which the effective interfacial mobility is unity (Appendix~\ref{app:interface_units}),
\begin{equation}
j(s) = -\frac{\mathrm{d}\mu}{\mathrm{d}s} + j_A(\theta(s)),
\label{eq:edgecurrent}
\end{equation}
where $-\mathrm{d}\mu/\mathrm{d}s$ is the passive diffusive contribution and $j_A(\theta)$ is the angle-dependent active edge current.
While both of these terms typically depend on $s$, stationarity requires $j(s)= j$, where $j$ is a constant. Using Eq.~\eqref{eq:GibbsThomson} in Eq.~\eqref{eq:edgecurrent} then yields the equation
\begin{equation}
j = -\beta\,\frac{\mathrm{d}^2\theta}{\mathrm{d}s^2} + j_A(\theta).
\label{eq:theta_ode_raw}
\end{equation}
By introducing an active interface potential $V_A(\theta)$ via
\begin{equation}
j_A(\theta) = -\frac{\mathrm{d}V_A}{\mathrm{d}\theta},
\label{eq:VA_def}
\end{equation}
we can rewrite Eq.~\eqref{eq:theta_ode_raw} as
\begin{equation}
\frac{\mathrm{d}^2\theta}{\mathrm{d}s^2}
=
-\frac{\mathrm{d}}{\mathrm{d}\theta}\,V_{\rm eff}(\theta),
\qquad
\label{eq:Newton_mapping}
\end{equation}
where
\begin{equation}
    V_{\rm eff}(\theta)\equiv \frac{1}{\beta}\bigl[V_A(\theta)+j\,\theta\bigr]
    \label{eq:Veff_def}
\end{equation}
is defined as the effective interface potential.
Thus, as one moves along the interface, the orientation $\theta(s)$ follows the trajectory of a Newtonian particle
(with $s$ mapping to ``time'' and $\theta$ to ``position'') in the effective potential $V_{\rm eff}(\theta)$, represented by a gray ball in Fig.~\ref{fig:newton}b.
Accordingly, Eq.~\eqref{eq:Newton_mapping} admits the first integral
\begin{equation}
\frac{1}{2}\left(\frac{\mathrm{d}\theta}{\mathrm{d}s}\right)^2 + V_{\rm eff}(\theta) = \mathcal{E},
\label{eq:energy_integral}
\end{equation}
which can be interpreted as an effective total energy that controls the ``velocity'' $\mathrm{d}\theta/\mathrm{d}s$ at any given $\theta$ and thus the total length of the contour. Once $\theta(s)$ is determined, the condensate contour follows by integrating ${\bf t}(s)$.

A physically admissible condensate requires a closed interface contour, such that 
\begin{equation}
    \oint \mathbf t(s)\,\mathrm{d}s
    =  \mathbf 0.
    \label{eq:spatial_closure}
\end{equation}
In the mapping to a Newtonian particle, this requires that the particle trajectories return to the same point in phase space with the same velocity after a $2\pi$ advance. 

The regular $n$-sided polygons ($n\geq3$) observed in Fig.~\ref{fig:camb}c--f
result from an effective potential with period $2\pi/n$,
i.e., $V_{\rm eff}(\theta+2\pi/n)=V_{\rm eff}(\theta)$, which originates from the $n$-fold symmetry of the active edge current $j_A(\theta)$.
The condensate contour then consists of $n$ identical segments, successively rotated by $2\pi/n$, which guarantees
closure.
Such a periodic potential requires that $j$ takes a particular value $j=j^*$, with
\begin{equation}
j^\ast
\equiv
\frac{1}{2\pi}\int_{0}^{2\pi}\mathrm{d}\theta\, j_A(\theta).
\label{eq:j_star}
\end{equation}
When $j=j^*$, the effective potential $V_\mathrm{eff}(\theta)$ is periodic and exhibits maxima for values $\theta=\theta^*$, where the interface curvature $\mathrm{d}\theta/\mathrm{d}s$ is minimal.    
If $\mathcal{E}$ is only slightly larger than these maxima, the trajectory $\theta(s)$ spends very long ``time'' there, corresponding to long straight sections of the interface with constant angles $\theta^*$, the facet angle.
The potential landscape thus determines the set of facet angles, while the conserved quantity
$\mathcal{E}$ sets the overall size of the condensate and the relative lengths of
facets. If the area fraction of the condensate is fixed, $\mathcal{E}$ needs to be chosen to satisfy this global constraint.

As a 4-fold-symmetric example we choose
\begin{equation}
j_A(\theta)=\zeta\bigl(j_0-\cos 4\theta\bigr),
\label{eq:jA_4fold}
\end{equation}
with activity strength $\zeta$ and offset $j_0$, similar to what we chose in Eq.~\eqref{eq:cambja}. Then, Eq.~\eqref{eq:j_star} gives $j^\ast=\zeta j_0$ and Eq.~\eqref{eq:Veff_def} with $j=j^\ast$ yields a purely periodic
effective potential
\begin{equation}
V_{\rm eff}(\theta)
=
\frac{\zeta}{4\beta}\,\sin(4\theta) ,
\label{eq:Veff_4fold}
\end{equation}
whose maxima select, for $\zeta>0$, the stable facet orientations
$\theta^\ast=\pi/8 + m\pi/2$ ($m\in\mathbb{Z}$), corresponding to a tilted-square condensate with rounded corners.
For $j\neq j^\ast$, $V_{\rm eff}$ acquires a linear tilt and the resulting interfacial trajectories do not close, corresponding to physically impossible condensate shapes (Fig.~\ref{fig:newton}b-d).

Given a measured current--angle relation $j_A(\theta)$ from the chiral active lattice gas simulations, we construct
$V_A(\theta)=-\int^\theta \mathrm{d}\theta'\, j_A(\theta')$ (up to an irrelevant constant), determine $j^\ast$ from
Eq.~\eqref{eq:j_star}, and obtain $V_{\rm eff}(\theta)$ from Eq.~\eqref{eq:Veff_def}.
The values of $\theta$ for which the resulting $V_{\rm eff}$ is maximal correspond to the stationary facet angles $\theta^* \approx 0.384$, in quantitative agreement with the interface orientations observed in simulations at $p_c = 1$ (Fig.~\ref{fig:newton}e). Thus, measuring the angular dependence of the edge current is sufficient to predict the facet angles.

\section{Discussion}

We introduced a minimal model of a chiral active lattice gas in two dimensions with conserved particle number. Building on the nearest-neighbor interactions of the Ising model, chirality and activity are implemented simultaneously through stochastic biased local rotations. 
At low temperature, the system undergoes phase separation into chiral condensates with interfaces that are faceted and tilted at characteristic angles with respect to the lattice axes. 
At these interfaces, persistent, unidirectional, angle-dependent edge currents emerge from the local chiral active dynamics. 
Inspired by these observations, and to discuss generic features, we extended Model-B dynamics by adding an active edge current term to the Cahn--Hilliard model. 
We show that an $n$-fold symmetric active edge current reshapes a circular condensate into a shape resembling a regular $n$-sided polygon, while preserving the bulk densities and interfacial profile of the equilibrium model. 
In the thin-interface limit, we define an effective interface potential and construct a dynamic principle that uniquely determines the geometry of the condensate contour, in agreement with the chiral active lattice gas simulations.

Chiral activity can be introduced into lattice gases in several ways. The minimality of our approach is ensured by two features. 
The $2\times2$ box rotation is the smallest local unit capable of realizing a rotation bias, and the ergodicity of the resulting dynamics guarantees that all nonequilibrium steady states are dynamically accessible from arbitrary initial conditions.
Extensions to other rotation assemblies or combinations with Kawasaki exchange dynamics may provide further minimal routes to chiral activity if ergodicity is preserved. 
At the continuum level, the active edge current term we introduce is chiral and cannot be expressed as the derivative of a free energy. 
Incorporating this term into active Model~B+~\cite{tjhung2018cluster} would provide a more complete active field model of phase separation that includes the interplay of self-propulsion and chirality.  
An interesting open question is whether this interplay can produce faceted analogues of bubbly phase separation. 
More broadly, we reveal the microscopic origin of edge currents in this chiral active lattice gas: interface excitations are created in pairs of opposite type and propagate in opposite directions along the interface, phenomenologically reminiscent of electron-hole pairs in semiconductors~\cite{kittel2018introduction}. 
Boundary-localized unidirectional transport is also a general feature shared with other systems like quantum Hall systems~\cite{halperin1982quantized,buttiker1988absence}. Given the generality of the Ising model beyond classical physics, it would be interesting to explore quantum versions of this chiral active lattice gas, to test how quantum coherence and statistics may modify the resulting observations.

\section*{Data Availability}
The simulation code used to generate the results of this article is openly available on Zenodo~\cite{wang2026code}.

\appendix
\setcounter{figure}{0}
\renewcommand{\thefigure}{A\arabic{figure}}

\section{Proof of Ergodicity}
\label{app:ergodicity}

For lattice gases, ergodicity means that any two configurations can be connected through a finite sequence of local updates. 
For our dynamics, this means that it must be possible to exchange any pair of opposite-spin sites without affecting the rest of the system, by decomposing the global transformation into a sequence of allowed local rotations. 
We should also point out that, even when $p_c=1$, three clockwise rotations on the same box are equivalent to a single counterclockwise rotation.  
Conversely, when $p_c=0$, three counterclockwise rotations are equivalent to a single clockwise rotation. Hence, any proof involving rotations in arbitrary direction also applies to $p_c=1$ and $p_c=0$.

\begin{figure}
  \centering
  \includegraphics[width=.47\textwidth]{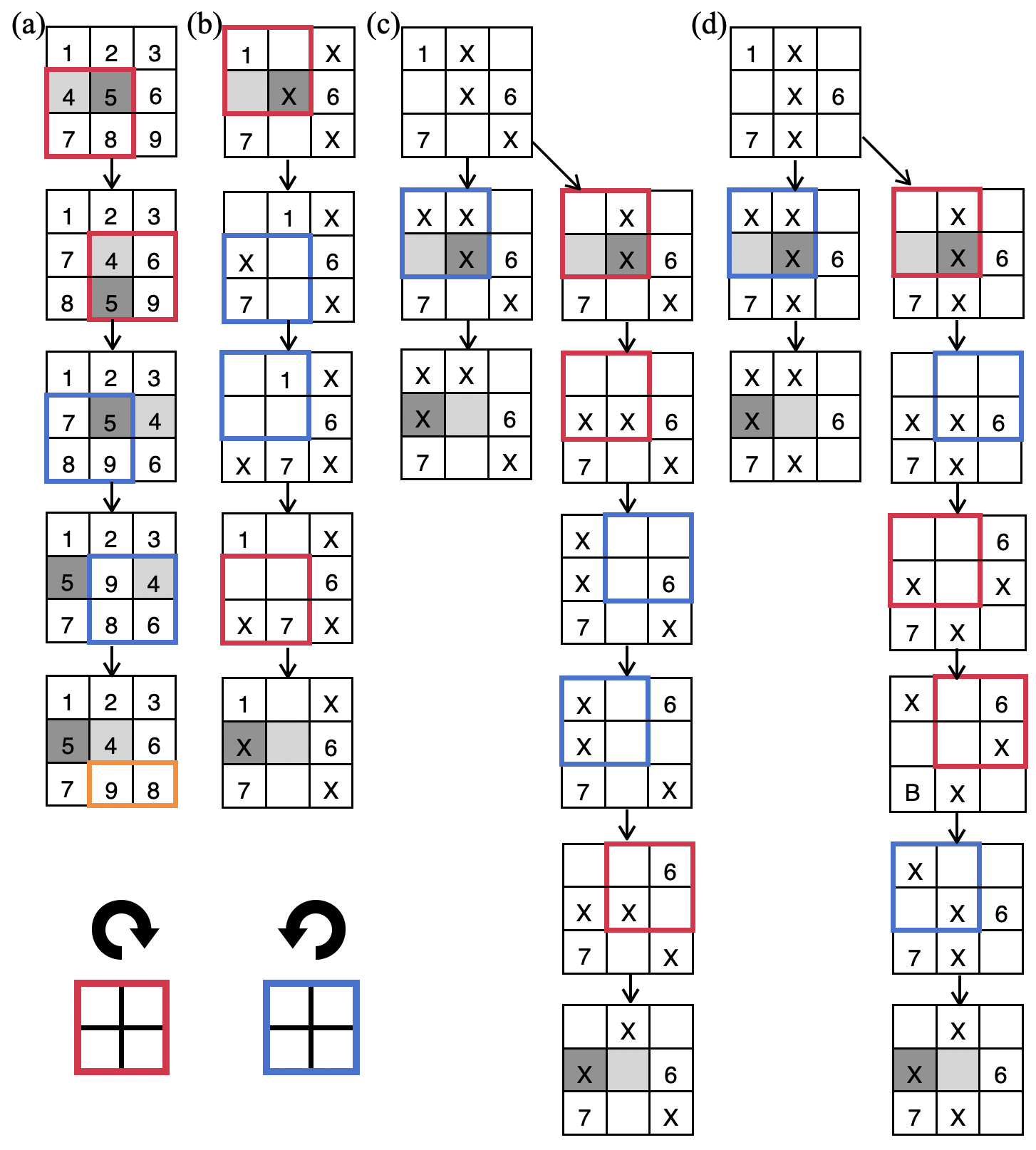}
  \caption{Illustration for the proof of ergodicity, showing how the states of sites labeled 4 and 5 at the top of column (a) can be exchanged through local rotations (clockwise: red, counterclockwise: blue), without affecting other sites. The strategy of column (a) leads to the exchange of sites 8 and 9, which only matters if they are in different states. The remaining columns (b-d) show alternative strategies for the case where both sites pair 2-3 and 8-9 are composed of different states, labeled as empty and ``X'' (which can stand for type~A and B particles, or vice versa). For those strategies, also the states of some other neighboring sites need to be specified.}
  \label{fig:ergodicity}
\end{figure}

To establish the above spin swap condition, we first focus on a local $3 \times 3$ block, where the nine sites are labeled from 1 to 9, with site 5 at the center (Fig.~\ref{fig:ergodicity}(a)). Due to the four-fold rotational symmetry of the square lattice, it suffices to show that we can swap the spins of site 5 and any of its nearest neighbors--for example, site 4. If this local operation is possible, any nearest-neighbor-pair swap can be performed elsewhere by symmetry.
We demonstrate that sites 4 and 5 can be exchanged by following a specific sequence of local rotations shown in Fig.~\ref{fig:ergodicity}(a), which involves sacrificing the 8--9 spin pair, because they are swapped. By symmetry, we can also choose to sacrifice the 2--3 spin pair. If either the 2--3 or 8--9 pair consists of identical spins, this operation could leave the rest of the configuration unchanged.

Next, we address the remaining special cases, where neither 2--3 nor 8--9 consists of identical spins. 
Due to the spin-inversion symmetry of the system, we only need to consider one representative case where in Fig.~\ref{fig:ergodicity}(a) site 4 is spin-down (empty) and site 5 is spin-up (denoted by ``X''). 
Under this setup, the only three cases are shown at the top of Fig.~\ref{fig:ergodicity}(b)--(d). In each configuration, we still keep other sites labeled as ``1,'' ``6'', and ``7'' to indicate that their spin values are arbitrary and do not affect the result. 
In each case, the following sequence of rotations guarantees that only the spins at sites 4 and 5 are exchanged.
Therefore, all configurations relevant to nearest-neighbor spin exchanges have been considered, which in turn generates all sequential non-neighbor exchanges. This confirms that our local dynamical rule is sufficient to ensure ergodicity.

\section{Quantitative analysis of lattice patterns}
\label{app:detail}

We set simulation temperature to $k_BT/J=1.2$, corresponding to $T\simeq0.53T_c$ of the equilibrium square-lattice Ising model~\cite{onsager1944crystal}. At this temperature, phase segregation is pronounced and biased rotations sustain unidirectional edge currents, while thermal fluctuations keep lattice-induced orientational bias weak. Fourfold condensate with preferred tilt angles and edge currents can also emerge at other temperatures satisfying these conditions.

To compare orientational anisotropy under biased rotations, unbiased rotations, and equilibrium Kawasaki dynamics, we define the coherent
fourfold component of the low-wavenumber structure factor as
\begin{equation}
C_4 =
\left|
\left\langle
\frac{\sum_{0<|\mathbf{k}|\leq 1}S(\mathbf{k})e^{4i\theta_{\mathbf{k}}}}
     {\sum_{0<|\mathbf{k}|\leq 1}S(\mathbf{k})}
\right\rangle
\right|,
\label{eq:C4}
\end{equation}
where $\theta_{\mathbf{k}}$ is the polar angle of $\mathbf{k}$ and the
average is taken over independent late-time configurations. At
$k_BT=1.2J$, we obtain
\begin{equation}
C_4 =
\begin{cases}
0.538\pm0.008, & p_c=0 \quad (N=32),\\
0.543\pm0.009, & p_c=1 \quad (N=32),\\
0.043\pm0.019, & p_c=1/2 \quad (N=6),\\
0.037\pm 0.012, & \mathrm{Kawasaki}\quad (N=6).
\end{cases}
\end{equation}
Here, $N$ is the number of independent runs, and uncertainties are bootstrap standard errors. The results for $p_c=0$ and $p_c=1$ agree within uncertainties, as expected from mirror symmetry, $C_4(p_c)=C_4(1-p_c)$. Both exhibit pronounced fourfold anisotropy, consistent with their nearly straight interfaces and cross-shaped Fourier spectra. In contrast, unbiased rotations and Kawasaki dynamics yield similarly small $C_4$, indicating weak orientational anisotropy on the domain scale in both equilibrium cases.

To assess finite-size effects, we compare $L=256$, $512$, and $1024$ at $k_BT=1.2J$, $p_c=1$, and type-A fraction $0.4$, using six independent runs from random initial conditions for each size. The $L=512$ data are runs 1--6 of the 32-run ensemble in the main text. From the two-dimensional Fourier spectra, we extract the preferred tilt angle as in Fig.~\ref{fig:simulation}:
\begin{align}
\theta^*(L=256)  &= 0.3857 \pm 0.0049, \notag\\
\theta^*(L=512)  &= 0.3845 \pm 0.0021, \\
\theta^*(L=1024) &= 0.3853 \pm 0.0004. \notag
\end{align}
The uncertainties are standard errors over the six runs. The preferred angles agree within uncertainties, with no resolved systematic size dependence, although smaller systems show larger run-to-run fluctuations.

\section{Effective  edge current}
\label{app:edge_current}

\begin{figure}[t]
  \centering

  \includegraphics{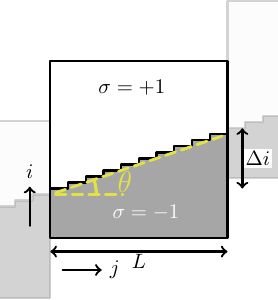}

  \caption{
  Schematic of the skewed periodic boundary condition. The central
  box denotes the simulation field, while the lighter boxes
  show parts of its neighboring periodic images. Translation by one
  system length $L$ in the $j$ direction is accompanied by an integer
  displacement $\Delta i$ in the $i$ direction, such that
  $\sigma_{(i+\Delta i,j+L)}=\sigma_{(i,j)}$. The staircase-like
  interface is therefore continuous across the identified boundaries
  and has nominal tilt angle $\theta$. Spins outside the
  two non-periodic boundaries in the $i$ direction are fixed to
  opposite phases.
  }
  \label{fig:skewed_boundary}

\end{figure}

We measure the  edge current in Monte Carlo simulations of our lattice model as follows.
We run simulations with skewed periodic boundary conditions enforcing a
series of prescribed interface tilt angles. Specifically, translation
by one system length $L$ in the $j$ direction is accompanied by an
integer displacement $\Delta i$ in the $i$ direction, such that
\begin{equation}
\sigma_{(i+\Delta i,j+L)}
=
\sigma_{(i,j)}.
\end{equation}
The displacement $\Delta i$ determines the nominal interface tilt angle
according to
\begin{equation}
\tan\theta
=
\frac{\Delta i}{L}.
\end{equation}
The spins outside the two non-periodic boundaries in the $i$ direction
are fixed to opposite phases, with $\sigma=-1$ for $i<0$ and
$\sigma=+1$ for $i\geq L$. These boundary conditions maintain a single
tilted interface. The geometry is illustrated in
Fig.~\ref{fig:skewed_boundary}.

The local current at lattice site
$\mathbf{m}$ is defined as the vector formed by the net number of
type-A particles crossing the rightward and upward nearest-neighbor boundaries
per Monte Carlo sweep.
In Monte Carlo sweep $s$, the net horizontal and vertical transport of
type-A particles across the rightward and upward nearest-neighbor boundaries
adjacent to lattice site $\mathbf{m}$ are denoted by
$h_\mathbf{m}^s$ and $v_\mathbf{m}^s$, respectively, where
\begin{equation}
h_\mathbf{m}^s = A_{\rightarrow} - A_{\leftarrow},
\qquad
v_\mathbf{m}^s = A_{\uparrow} - A_{\downarrow},
\end{equation}
and $A_{\dots}$ denotes the total number of type-A particles moved in the
direction indicated by the subscript arrow during sweep $s$.

Averaging over the first $R$ Monte Carlo sweeps, we obtain the
time-averaged local current components
\begin{equation}
\bar{J}_{h}^\mathbf{m} = \frac{1}{R} \sum_{s=1}^{R} h_\mathbf{m}^s,
\qquad
\bar{J}_{v}^\mathbf{m} = \frac{1}{R} \sum_{s=1}^{R} v_\mathbf{m}^s .
\end{equation}
The time-averaged local current vector at lattice site $\mathbf{m}$ is
then given by
\begin{equation}
\mathbf{J}^{\,\mathbf{m}}
=
\left(
\bar{J}_{h}^\mathbf{m},
\bar{J}_{v}^\mathbf{m}
\right).
\end{equation}

\begin{figure}
  \centering
  \includegraphics[width=.33\textwidth]{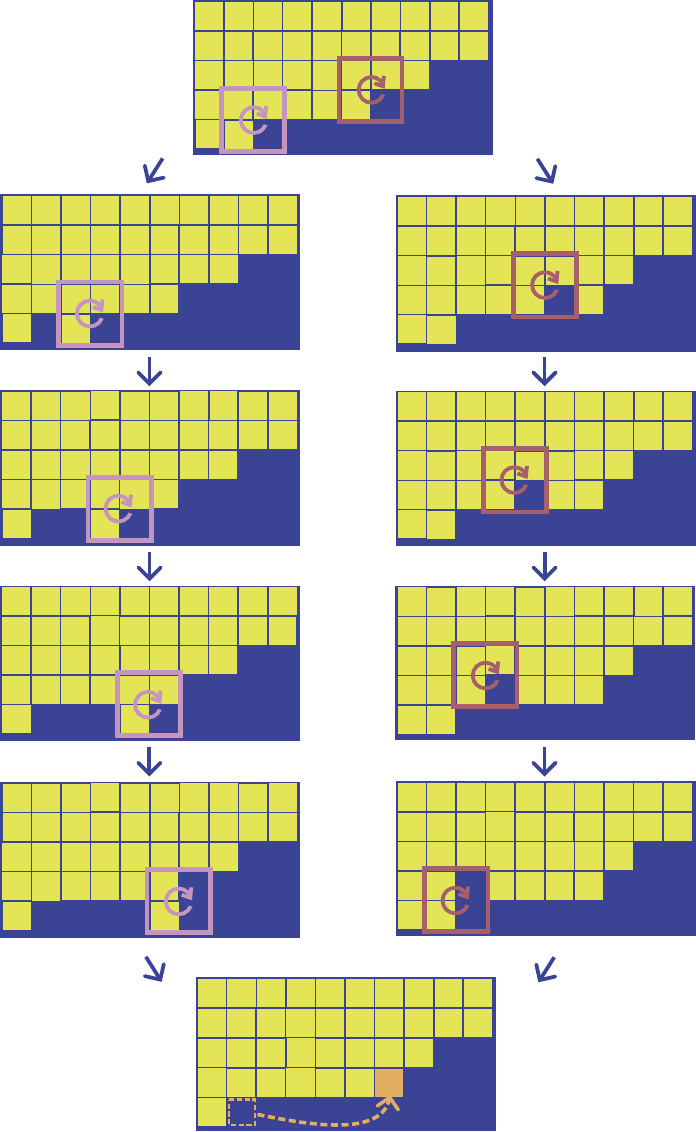}
  \caption{
  Elementary moves along lattice interfaces under $p_c=1$ dynamics in the limit of small $T$.
 Sequence of local rotation events illustrating how a type-A particle located at a corner site of a tilted interface is transported stepwise along the boundary. Only particles at corner positions contribute to the net tangential flux, while bulk particles remain immobile at low temperature. Repeated rotations generate a persistent edge current localized at the interface.
 At low temperatures, the rotations indicated in the top panel are much energetically less favored than the fast moves they are immediately followed by, shown in the remaining panels. The left column shows a transport of a particle of the yellow phase, the right column the transport of a hole. Both sequences result in the same final state (bottom panel), having transported one particle from the end of one row of the yellow phase to the end of the row above. }
  \label{fig:element}
\end{figure}

At low temperature, flux events are observed to be localized at the
interface (Fig.~\ref{fig:element}) and approximately uniformly
distributed along the interface, provided that the interface remains essentially
undeformed. The net  edge current $j(\theta)$ for an interface with tilt
angle $\theta$ is therefore estimated as the magnitude of the spatially
averaged current vector,
\begin{equation}
j(\theta)
=
\left|
\frac{\cos\theta}{L}
\sum_{\mathbf{m}}
\mathbf{J}^{\,\mathbf{m}}
\right|.
\end{equation}
Here, the summation runs over the entire lattice; since
$\mathbf{J}^{\,\mathbf{m}} \approx 0$ in the bulk, the sum
effectively receives contributions only from lattice sites at the
interface. The factor $\cos\theta$ accounts for the geometric projection
between the interface length and the square lattice edge length $L$.

For the data shown in Fig.~\ref{fig:simulation}f, we take $L=512$ and $R=20$, which is sufficient for the net flux to be
generated uniformly along the interface, while the interface remains
essentially undeformed, even for slopes that are unstable at longer
times.

\section{Numerical protocol for the chiral active Model~B}
\label{app:camb_numerics}

To solve Eq.~\eqref{eq:camb} numerically, we divide the simulation domain into a uniform square grid of $128\times128$ boxes of side length $\Delta x=\Delta y=1$, with periodic boundary conditions in both directions. The continuous field $\phi(\mathbf r,t)$ is represented by its values $\phi_{i,j}(t)$ at the box centers $\mathbf r_{i,j}=(i\Delta x,\,j\Delta y)$, and currents are defined on the faces shared by neighboring boxes; for example, $J^x_{i+1/2,j}$ denotes the current from box $(i,j)$ into box $(i+1,j)$.

In each time step, we perform three operations on every box. First, we compute the chemical potential of Eq.~\eqref{eq:mu},
\begin{equation}
\mu_{i,j}=a\phi_{i,j}+b\phi_{i,j}^3-K\,(\nabla^2\phi)_{i,j},
\end{equation}
where the discrete Laplacian is given by the isotropic nine-point stencil
\begin{equation}
(\nabla^2\phi)_{i,j}
=
\frac{1}{6\Delta x^2}
\Big[
4\!\sum_{\text{nn}}\phi
+\!\sum_{\text{diag}}\phi
-20\,\phi_{i,j}
\Big],
\end{equation}
with the first sum running over the four nearest neighbors of box $(i,j)$ and the second over its four diagonal neighbors. Second, we compute the total current $\mathbf J=-M\nabla\mu+\mathbf J_A$ of Eq.~\eqref{eq:camb} on each face of the box. Using $\boldsymbol\varepsilon\cdot\nabla\phi=(\partial_y\phi,\,-\partial_x\phi)$ in Eq.~\eqref{eq:JA}, the currents through the right and upper faces are
\begin{equation}
\begin{aligned}
J^x_{i+1/2,j}&=-M\,\frac{\mu_{i+1,j}-\mu_{i,j}}{\Delta x}
+j_A(\theta)\,\partial_y\phi\,\Big|_{i+1/2,j},\\
J^y_{i,j+1/2}&=-M\,\frac{\mu_{i,j+1}-\mu_{i,j}}{\Delta y}
-j_A(\theta)\,\partial_x\phi\,\Big|_{i,j+1/2},
\end{aligned}
\end{equation}
where $\nabla\phi$ on each face is evaluated by finite differences of $\phi$ in the neighboring boxes, and $\theta$ follows from its direction via Eq.~\eqref{eq:theta}. Third, we update the field according to the first line of Eq.~\eqref{eq:camb}, using an explicit Euler step of size $\delta t$:
\begin{equation}
\phi_{i,j}(t+\delta t)
=\phi_{i,j}(t)
-\frac{\delta t}{\Delta x}\left(\delta_x J^x+\delta_y J^y\right)_{i,j},
\end{equation}
where
\begin{equation}
\begin{aligned}
(\delta_x J^x)_{i,j}&=J^x_{i+1/2,j}-J^x_{i-1/2,j},\\
(\delta_y J^y)_{i,j}&=J^y_{i,j+1/2}-J^y_{i,j-1/2}
\end{aligned}
\end{equation}
are the differences between the currents through opposite faces of box $(i,j)$.

Unless otherwise stated, we use the parameters $a=-1/4$, $b=1/4$, and $K=M=1$, and the edge current of Eq.~\eqref{eq:cambja} with $j_0=3/2$ and $|\zeta|=0.1$; the symmetry order $n$ and the sign of $\zeta$ are given in the figure captions. We use the time step $\delta t=10^{-4}$ and integrate up to $t=8000$, saving the field every $800$ time units.

We use two types of initial conditions, both with spatial mean $\bar\phi\simeq-0.31$. For the random initial condition (Fig.~\ref{fig:camb}b), each box is independently set to $\phi_{i,j}=+1$ or $-1$ with probabilities $0.3$ and $0.7$, respectively. For the circular initial condition (Figs.~\ref{fig:camb}c--f), we instead set $\phi_{i,j}=+1$ inside a circle centered in the box and covering $30\%$ of its area, and $\phi_{i,j}=-1$ outside. In both cases, we then add to every $\phi_{i,j}$ an independent random number drawn uniformly from $[0.08,\,0.10]$.

\section{Interface profile}
\label{app:profile}

The center-line of the interface in chiral active Model~B is defined by the contour $\phi=0$. We measure the $\phi$ profile as a function of the normal distance from the interface, $\xi$, in units of one grid spacing.

\begin{figure}[t]
\includegraphics[width=0.8\linewidth]{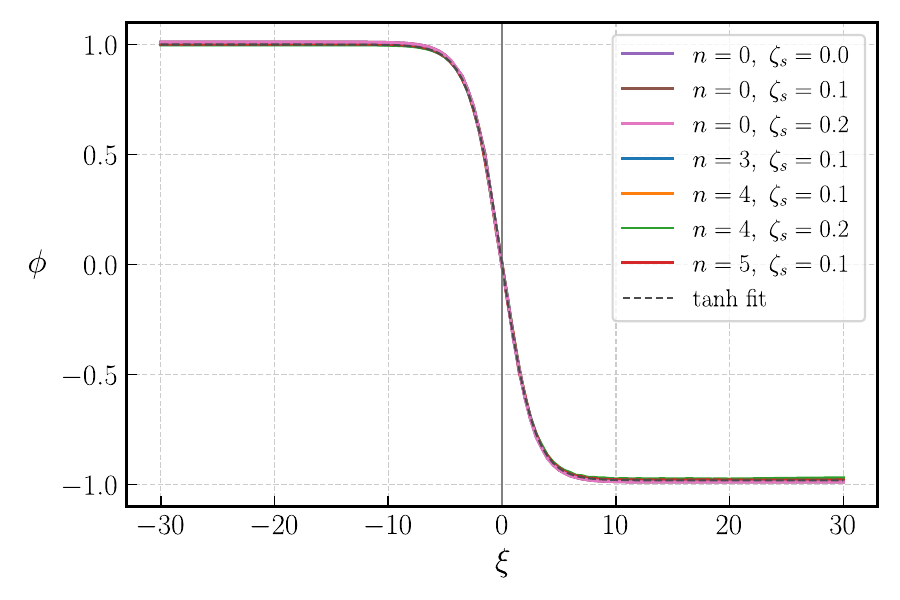} 
\caption{Spatially averaged field profiles $\phi(\xi)$ for each $(n,\zeta)$, shown as a function of the signed normal distance $\xi$ from the interface ($\phi=0$ contour). A representative tanh fit, $\phi(\xi)=A\tanh[(\xi-\xi_0)/w]+\phi_0$, is overlaid as a dashed line.}
\label{fig:profile}
\end{figure}

To assign a signed distance to every grid $\mathbf{r}$, we compute the shortest
Euclidean distance $d(\mathbf{r})$ to the interface contour and define
\begin{equation}
\xi(\mathbf{r}) =
\begin{cases}
-\,d(\mathbf{r}), & \mathbf{r} \text{ inside },\\
\phantom{-}d(\mathbf{r}), & \mathbf{r} \text{ outside}.
\end{cases}
\end{equation}
We bin all points by $\xi$ with spacing $\Delta\xi=0.5$, within the window
$|\xi|\le 35$, and compute the mean profile in each bin:
\begin{equation}
\bar{\phi}(\xi_k)=\frac{1}{N_k}\sum_{\mathbf{r}\in k}\phi(\mathbf{r}),
\end{equation}
where $N_k$ is the number of grids in bin $k$. We then fit the averaged profile to the function
\begin{equation}
\phi(\xi)=A\,\tanh\!\left(\frac{\xi-\xi_0}{w}\right)+\phi_0,
\end{equation}
where $A$ is the amplitude, $\xi_0$ is the interface location, $w$ is the interfacial width.
As shown in Fig.~\ref{fig:profile}, increasing the strength or fold of symmetry of the active  edge current will not distinctively change either the interface profile or the bulk density from the equilibrium circular condensate($n=0, \zeta=0$). 

\begin{figure}[t]
    \centering
    \includegraphics[width=0.63\linewidth]{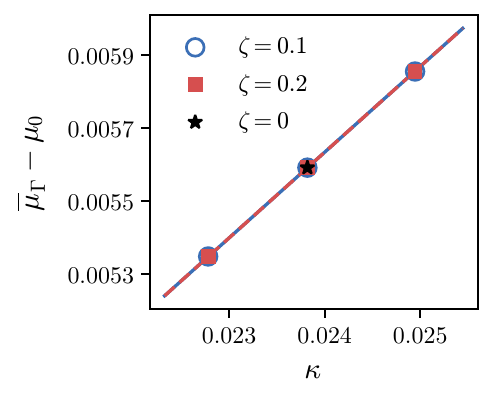}
    \caption{The interface-averaged chemical-potential difference
  $\overline{\mu}_{\Gamma}-\mu_0$ as a function of the mean curvature
  $\kappa$ for circular condensates at $\bar\phi=0.27$, $0.30$, and $0.33$, with
  $\zeta=0.1$ and $0.2$. The $\zeta=0$ control point is evaluated at $\bar\phi=0.30$.}
    \label{fig:circles}
\end{figure}

The active edge current does not enter the definition of the chemical
potential $\mu$. We therefore test the curvature contribution using
initially circular condensates with an angle-independent active edge current
and different packing fractions, with the same model parameters
and numerical protocol as above. For each final state, $\Gamma$ denotes the $\phi=0$
contour, $\kappa$ its mean curvature, and
$\overline{\mu}_{\Gamma}$ the interface-averaged chemical potential. 
Figure~\ref{fig:circles} shows that $\overline{\mu}_{\Gamma}-\mu_0$ 
remains proportional to $\kappa$ with a common fitted slope.

\FloatBarrier
\section{Stability analysis of steady-state interfaces in chiral active Model~B}
\label{app:stability}
We consider a straight section of the interface with a steady slope $\tan\theta^*$. The coordinates $(x,y)$ of the points on the interface are given by $y = x\tan \theta^*$. Now we set $s$ as the tangential direction along the interface and $n$ as the normal direction. 
We then use this straight interface as a reference for a new coordinate system $(s,n)$, where
\begin{align}
\begin{cases}
s &= x \cos \theta^* + y \sin \theta^*,\\
n &= -x \sin \theta^* + y \cos \theta^*.
\end{cases}
\label{eq:sn}
\end{align}
For a perturbed, dynamically evolving interface, we denote the height in normal direction as $h(s,t)$, where $t$ is time. In the thin interface limit, $n>h(s,t)$ is the inside condensate phase (+) and $n<h(s,t)$ is the outside phase (--).

With the thin interface limit, the normal velocity of the interface $\partial_t h(s)$ is proportional to the difference of input and output fluxes in both normal and tangential directions\cite{bray1994theory,zwicker2017growth}, which is governed by
\begin{equation}
\partial_t h(s) = \left.\frac{J_n^+(s,n) - J_n^-(s,n) + \partial_sJ_s(s)}{\phi^+(s,n) - \phi^-(s,n)}\right|_{n=h(s)},
\label{eq:stefan}
\end{equation}
where $\phi^{\pm}(s,n)$ are the corresponding bulk densities,  $J_n^\pm(s,n)$ are the diffusion fluxes in normal direction from the inside and the outside of the condensate, which, in the long-wavelength limit, take the form
\begin{equation}
J_n^\pm(s,n) = -D\partial_n \phi^\pm(s,n),
\label{eq:jn}
\end{equation}
where $D$ is the effective diffusivity. Compared to equilibrium systems, chiral active Model~B system has an extra term $\partial_s J_s(s)$, which is the local edge current gradient along the interface. 

Now we apply a small perturbation to the interface  as a Fourier mode of wavenumber $q$ along the interface $s$, such that
\begin{equation}
h(s)= \hat{h} e^{iqs+\sigma t},
\end{equation}
with the initial amplitude $\hat h$ of a perturbation with wavenumber $q$, growing at the exponential rate $\sigma$,
and
\begin{equation}
\phi^{\pm}(s,n)= \phi^{\pm}_0 + \hat{\phi}^\pm(n) e^{iqs+\sigma t},
\label{eq:phiansatz}
\end{equation}
where $\phi^{\pm}_0=\pm\sqrt{-a/b}$ are the steady bulk densities, $\hat{\phi}^\pm(n)$ are functions with infinitesimal amplitude, and $\sigma$ is the growth rate. 
The scalar field satisfies the diffusion equation
\begin{equation}
\frac{\partial \phi^{\pm}}{\partial t} = D \nabla^2 \phi^{\pm},
\label{eq:effdiffusion}
\end{equation}
where $D=-2aM>0$ is obtained by
linearizing the passive chemical potential
about $\phi_0^\pm$
and neglecting higher-order spatial gradients in the
long-wavelength limit~\cite{bray1994theory}.

For the constant bulk densities, we have $\partial_t \phi^{\pm}_0 = 0$ and $\nabla^2 \phi^{\pm}_0 = 0$, such that plugging \eqref{eq:phiansatz} into \eqref{eq:effdiffusion} leads to
\begin{equation}
    \sigma \hat{\phi}^\pm = D\left(-q^2+\frac{\partial^2}{\partial n^2}\right)\hat{\phi}^\pm.
\end{equation}
The bounded solutions are
\begin{equation}
\hat{\phi}^+(n) = A^+ e^{-\lambda n}, \quad \hat{\phi}^-(n) = A^- e^{\lambda n},
\end{equation}
where $\lambda = \sqrt{q^2 + \sigma/D}$. The Gibbs--Thomson boundary condition in Eq.~\eqref{eq:GibbsThomson} relates the interface curvature $\kappa=\partial_s^2 h$ to the chemical potential $\mu$ at the interface, which in turn determines the boundary values of $\phi^\pm$ at the interface. Since $1/\lambda\gg h(s)$ holds for small perturbations, we get
\begin{equation}
    A^+= A^- = -\nu q^2 \hat{h},
\end{equation}
where $\nu=\beta/(-2a)>0$ is an effective capillary coefficient, obtained by the linearized bulk relation \mbox{$\delta\mu=-2a\,\delta\phi$}. Therefore, under a small perturbation,
\begin{equation}
    \partial_n \hat \phi^+|_{n=h(s)} = \nu q^2 \hat{h}\lambda, \quad  
    \partial_n \hat \phi^-|_{n=h(s)} = -\nu q^2 \hat{h}\lambda.
\end{equation}

Now, we consider the tangential flux, where $j_A$ is the only contribution, such that  
\begin{equation}
    J_s = \Delta\phi\, j_A(\theta),
\end{equation}
where $\theta(s)$ is the local slope angle and $\Delta\phi=\phi^{+}-\phi^{-}$.
At time $t=0$, the interface height is $h(s,t=0) = \hat h e^{iqs}$ and the perturbed local slope angle is
\begin{equation}
\theta(s) = \theta^* + i q \hat h e^{iqs},
\end{equation}
where \( \theta^* \) is the steady slope angle with respect to the original coordinate. 

For a small perturbation, we expand the nonlinear term using Taylor expansion:
\begin{align}
    j_A(\theta (s)) 
    &= j_A\left( \theta^* + i  q \hat h e^{iqs} \right) \notag\\
    &= j_A(\theta^*)+ j_A'(\theta^*)i q \hat h  e^{iqs}
    + \mathcal{O}(\hat h^2),
\end{align}
where $j_A'$ denotes the derivative with respect to $\theta$. The tangential derivative of the  edge current then becomes
\begin{align}
    \partial_s J_s 
    &= \Delta\phi\, \partial_s j_A(\theta(s)) \notag\\
    &=  - q^2 \Delta\phi\, j_A'(\theta^*)\, \hat h e^{iqs}
    + \mathcal{O}(\hat h^2).
\end{align}
Using Eq.~\eqref{eq:stefan} and Eq.~\eqref{eq:jn}, we obtain
\begin{equation}
\sigma=-\frac{2D\nu q^{2}}{\Delta\phi}\sqrt{q^{2}+\frac{\sigma}{D}}
-q^{2}\,j_A'(\theta^{*}).
\label{eq:disp}
\end{equation}

In the long-wavelength limit $q\to 0$, the dispersion relation is
dominated by the $q^{2}$ contribution from the tangential edge current,
\begin{equation}
\sigma(q)=-j_A'(\theta^{*})q^{2}+O(q^{3}).
\label{eq:longwave}
\end{equation}

Therefore, the linear stability of the straight interface at long
wavelengths is controlled by the sign of $j_A'(\theta^{*})$. The stability criterion is thus
\begin{align}
j_A'(\theta^{*})&>0 \;\Rightarrow\; \sigma(q)<0 \quad \text{(stable)},
\label{eq:stable}\\
j_A'(\theta^{*})&<0 \;\Rightarrow\; \sigma(q)>0 \quad \text{(unstable)}.
\label{eq:unstable}
\end{align}
This criterion is also consistent with our phenomenological stability argument for the chiral active Ising model in Fig.~\ref{fig:simulation}f.

By contrast, since $D>0$, $\nu>0$ and $\Delta\phi>0$, the first
term in Eq.~\eqref{eq:disp} stabilizes the interface within the thin interface limit $qw\ll1$, where $w$ is the interfacial width
defined in Appendix~\ref{app:profile}. When
$j_A'(\theta^*)<0$, setting $\sigma=0$ formally gives
\begin{equation}
q_c=\frac{|\Delta\phi\,j_A'(\theta^*)|}{2D\nu}.
\label{eq:qc}
\end{equation}
If $q_cw\ll1$, Eq.~\eqref{eq:disp} predicts growth for
$0<q<q_c$ and decay for $q_c<q\ll w^{-1}$.

When $qw=O(1)$, Eq.~\eqref{eq:disp} no longer determines the
growth rate. In the full continuum model, the square-gradient
contribution to the linearized Cahn--Hilliard dynamics provides
damping of order $-MKq^4$, which dominates the $O(q^2)$
edge-current contribution at sufficiently large $q$.
Thus the high-$q$ limit is stable, as in equilibrium
Cahn--Hilliard dynamics.

\vspace*{0.5\baselineskip}
\section{Nondimensionalization of edge currents}
\label{app:interface_units}

We specify the nondimensionalization that sets the effective mobility
to unity in Eq.~\eqref{eq:edgecurrent}. In the steady-state thin-interface
description, $j$ is the tangential current integrated across the
interface and normalized by the constant density jump $\Delta\phi$.
Integrating $\mathbf J_A=j_A(\theta)\,\boldsymbol\varepsilon\cdot\nabla\phi$
across the interface gives the tangential flux $\Delta\phi\,j_A(\theta)$
(see Appendix~\ref{app:stability}), so the active term enters $j$ as
$j_A(\theta)$. In dimensional form, the effective transport law reads
\begin{equation}
j(s)=-\mathcal M\,\partial_s\mu+j_A(\theta).
\label{eq:interface_units_dimensional}
\end{equation}
Here $\mathcal M>0$ is a constant effective interfacial mobility, treated
as a phenomenological parameter of the thin-interface description.
In the steady state,
interfacial mass conservation requires $j$ to be independent of $s$.

For fixed length and chemical-potential units $\ell_r$ and $\mu_r$,
we write $s=\ell_r\tilde s$, $\mu=\mu_r\tilde\mu$ and
$j=j_r\tilde j$, with the current unit $j_r=\ell_r^2/t_r$ set by the
time unit $t_r$, since $[j]=L^2/T$.
Substituting into Eq.~\eqref{eq:interface_units_dimensional} and
dividing by $j_r$ gives
\begin{equation}
\tilde j
=-\frac{\mathcal M\mu_r t_r}{\ell_r^3}\,
\frac{\mathrm d\tilde\mu}{\mathrm d\tilde s}
+\frac{j_A(\theta)}{j_r},
\end{equation}
where the prefactor of the first term is the dimensionless interfacial
mobility. Choosing
\begin{equation}
t_r=\frac{\ell_r^3}{\mathcal M\mu_r},
\qquad
j_r=\frac{\mathcal M\mu_r}{\ell_r},
\end{equation}
sets this prefactor to unity and yields Eq.~\eqref{eq:edgecurrent}.
In these units, $\zeta$ is expressed in units of $j_r$, since $j_A$ has
the same dimension as $j$, and $\beta$ in units of $\mu_r\ell_r$, since
$\mu-\mu_0=\beta\kappa$ with $[\kappa]=L^{-1}$.

\nocite{*}

\bibliography{bib}

\end{document}